\documentclass[sigconf]{acmart}
\AtBeginDocument{%
  }

\copyrightyear{2026}
\acmYear{2026}
\setcopyright{cc}
\setcctype{by}
\acmConference[ICSE-Companion '26]{2026 IEEE/ACM 48th International Conference on Software Engineering}{April 12--18, 2026}{Rio de Janeiro, Brazil}
\acmBooktitle{2026 IEEE/ACM 48th International Conference on Software Engineering (ICSE-Companion '26), April 12--18, 2026, Rio de Janeiro, Brazil}
\acmPrice{}
\acmDOI{10.1145/3774748.3793625}
\acmISBN{979-8-4007-2296-7/2026/04}

\acmISBN{978-1-4503-XXXX-X/2018/06}

\usepackage{amsmath,amsfonts}
\usepackage{algorithmic}
\usepackage{graphicx}
\usepackage{textcomp}
\usepackage{multirow}
\usepackage{booktabs}
\usepackage{soul}
\usepackage{ulem}
\usepackage{color}
\usepackage{xcolor}
\usepackage[ruled,linesnumbered]{algorithm2e}
\usepackage{url}
\usepackage{subcaption}
\usepackage{colortbl}
\usepackage{bbding}
\usepackage{listings}
\usepackage{cleveref}
\usepackage{enumitem}
\usepackage{pifont} 

\usepackage{float}
\usepackage{xspace}
\begin{document}

\title{LLM-Based Test Case Generation in DBMS through Monte Carlo Tree Search}


\author{Yujia Chen}
\affiliation{%
  \institution{Harbin Institute of Technology, Shenzhen}
  \city{Shenzhen}
  \country{China}
}
\email{yujiachen@stu.hit.edu.cn}

\author{Yingli Zhou}
\affiliation{%
  \institution{The Chinese University of Hong Kong, Shenzhen}
  \city{Shenzhen}
  \country{China}
}
\email{yinglizhou@link.cuhk.edu.cn}

\author{Fangyuan Zhang}
\affiliation{%
  \institution{Huawei Hong Kong Research Center}
  \city{Hongkong}
  \country{China}
}
\email{zhang.fangyuan@huawei.com}

\author{Cuiyun Gao}
\authornote{Corresponding author.}
\affiliation{%
 \institution{Harbin Institute of Technology, Shenzhen}
 \city{Shenzhen}
 \country{China}
}
\email{gaocuiyun@hit.edu.cn}






\begin{abstract}

Database Management Systems (DBMSs) are fundamental infrastructure for modern data-driven applications, where thorough testing with high-quality SQL test cases is essential for ensuring system reliability. Traditional approaches such as fuzzing can be effective for specific DBMSs, but adapting them to different proprietary dialects requires substantial manual effort. Large Language Models (LLMs) present promising opportunities for automated SQL test generation, but face critical challenges in industrial environments. First, lightweight models are widely used in organizations due to security and privacy constraints, but they struggle to generate syntactically valid queries for proprietary SQL dialects. Second, LLM-generated queries are often semantically similar and exercise only shallow execution paths, thereby quickly reaching a coverage plateau.

To address these challenges, we propose \textbf{MIST}, an LL\textbf{M}-based test case generat\textbf{I}on framework for DBM\textbf{S} through Monte Carlo \textbf{T}ree search. MIST consists of two stages: \textit{Feature-Guided Error-Driven Test Case Synthetization}, which constructs a hierarchical feature tree and uses error feedback to guide LLM generation, aiming to produce syntactically valid and semantically diverse queries for different DBMS dialects, and \textit{Monte Carlo Tree Search-Based Test Case Mutation}, which jointly optimizes seed query selection and mutation rule application guided by coverage feedback, aiming at boosting code coverage by exploring deeper execution paths. Experiments on three widely-used DBMSs with four lightweight LLMs show that MIST achieves average improvements of 43.3\% in line coverage, 32.3\% in function coverage, and 46.4\% in branch coverage compared to the baseline approach with the highest line coverage of 69.3\% in the Optimizer module.

\end{abstract}

\ccsdesc[500]{Computing methodologies~Artificial intelligence}
\ccsdesc[500]{Software and its engineering~Software post-development issues}

\keywords{DBMSs testing, Large language model, Test case generation}

\received[Received]{15 November 2025}
\received[revised]{6 January 2026}
\received[accepted]{6 January 2026}

\maketitle

\section{Introduction}

\begin{figure}[t]
    \centering
    \includegraphics[scale=0.46]{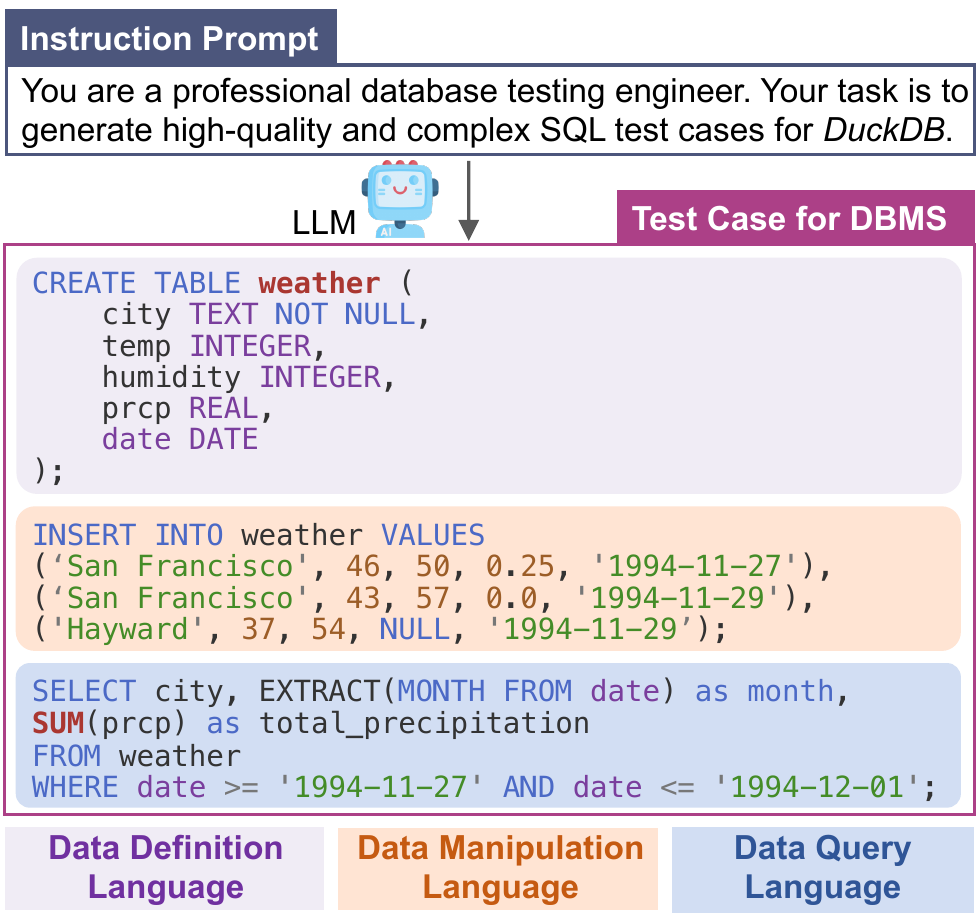}
    \caption{An illustrative example of using Qwen2.5-7B to generate a SQL test case for DBMS.}
    \label{fig:intro_case}
\end{figure}

\noindent Database Management Systems (DBMSs) are fundamental infrastructure for modern data-driven society, underpinning applications across finance, healthcare and e-commerce~\cite{stonebraker1986design,raasveldt2019duckdb,ozsu1996distributed}. Ensuring the correctness and robustness of DBMSs is vital, as even minor faults can lead to severe consequences, including data corruption, service outages, or security breaches~\cite{lo2010framework,chopra2010database,pick2025checking}. An essential step toward assuring DBMS reliability is thorough testing using high-quality SQL test cases, which help uncover subtle defects before deployment and prevent them from affecting production environments~\cite{zhong2020squirrel,bruno2006generating}. Traditional approaches for generating test cases, such as BuzzBee~\cite{yang2024towards}, SQLsmith~\cite{SQLsmith}, SQLancer~\cite{rigger2020testing}, can be effective for a specific DBMS; however, there exist many DBMSs designed for different application scenarios, such as DuckDB~\cite{raasveldt2019duckdb} for analytical workloads, PostgreSQL~\cite{stonebraker1986design} for general-purpose OLTP systems, and SQLite~\cite{owens2010sqlite} for lightweight embedded environments. Adapting these methods to different DBMSs requires substantial manual effort, such as crafting grammar rules and maintaining specific operators, which severely limits their scalability~\cite{zhong2025scaling,zhong2025testing,zhong2020squirrel}.
Recent advances in Large Language Models (LLMs) present promising opportunities for DBMS test case generation. First, LLMs have already demonstrated success in related data management tasks, including text-to-SQL~\cite{fan2024combining,li2024dawn,li2023can}, data cleaning~\cite{li2024autodcworkflow,naeem2024retclean,fan2024cost,qian2024unidm}, knob tuning~\cite{lao2023gptuner,giannakouris2025lambda}, and DBMS diagnosis~\cite{zhou2023d,zheng2024revolutionizing,singh2024panda}. Second, LLMs are applied in testing across multiple domains, such as general software testing~\cite{wang2024software,xue2024llm4fin}, compiler testing~\cite{gu2023llm,xie2025kitten}, and embedded testing~\cite{bley2025protocol}. As illustrated in Figure~\ref{fig:intro_case}, LLMs can directly generate SQL test cases by leveraging their understanding of SQL syntax and semantics acquired during pre-training. Despite these encouraging developments, deploying LLMs for industrial-scale DBMS testing still faces two critical challenges:

\begin{itemize}[leftmargin=*]

\item  \textbf{Challenge 1: Limited model adaptability in proprietary and resource-constrained industrial DBMS environments.} 
Industrial DBMSs often have proprietary SQL dialects with unique syntax and features. At the same time, security and privacy requirements restrict organizations to locally deployed lightweight LLMs (typically 70B parameters or smaller). The limited model capacity results in generating overly simple queries with insufficient testing coverage (see Figure~\ref{fig:challenges}(a)). Moreover, these models lack sufficient training on proprietary SQL variants, often producing queries that violate dialect-specific rules (see Figure~\ref{fig:challenges}(b)). As a result, adapting lightweight LLMs to proprietary industrial DBMSs remains challenging.

\item \textbf{Challenge 2: Limited exploration of deeper execution paths in test generation.}
Effective DBMS testing requires queries that cover meaningful execution paths beyond syntactic correctness. However, simply generating a large number of test cases does not translate into high coverage, as many queries are semantically similar and exercise only shallow execution paths. As demonstrated in a prior study~\cite{wang2021industry}, the generated test cases may improve coverage initially, but the coverage quickly plateaus as new inputs rarely explore deeper logic. Establishing an effective mechanism to guide test generation toward unexplored code regions remains a critical challenge.

\end{itemize}

\begin{figure}[t]
\begin{subfigure}[t]{0.38\textwidth}
\begin{minipage}{\linewidth}   
\begin{lstlisting}[language=SQL, basicstyle=\ttfamily, frame=single,  breaklines=true]
CREATE TABLE products (
    id INTEGER,
    name TEXT,
    price REAL
);
INSERT INTO products VALUES 
            (1, 'Laptop', 999.99),
            (2, 'Phone', 499.99);
SELECT name, price 
FROM products 
WHERE price > 10;
\end{lstlisting}
\end{minipage}
\caption{Simple query with limited testing coverage.}
\label{fig:challenge_semantic}
\end{subfigure}
\begin{subfigure}[t]{0.38\textwidth}
\begin{minipage}{\linewidth}
\begin{lstlisting}[language=SQL, basicstyle=\ttfamily, frame=single, escapechar=@]
CREATE TABLE employees (
  id INTEGER,
  details @\textcolor{red}{STRUCT<age: INTEGER>}@
);
INSERT INTO employees VALUES (1, 5);

SELECT * 
FROM products 
WHERE name @\textcolor{red}{REGEXP\_MATCHES}@('^A')
@\textcolor{red}{CROSS JOIN}@ cities c @\textcolor{red}{ON}@ c.id=1;
\end{lstlisting}
\end{minipage}
\caption{Query violating syntax rules on DuckDB (errors highlighted in \textcolor{red}{red}).}
\label{fig:challenge_syntax}
\end{subfigure}
\caption{Illustrating challenges in LLM-based DBMS test case generation with Qwen2.5-7B.}
\label{fig:challenges}
\end{figure}

\noindent \textbf{Our work.} To address these challenges, we propose \textbf{MIST}, an LL\textbf{M}-based test case generat\textbf{I}on framework for DBM\textbf{S} through Monte Carlo \textbf{T}ree search. MIST consists of two complementary stages: \textit{1) Feature-Guided Error-Driven Test Case Synthetization.} This stage constructs a feature tree from official documentation and hierarchically samples feature combinations to guide LLM generation. Execution errors are captured and fed back to subsequent prompts, improving syntactic validity and semantic diversity of synthesized test cases. \textit{2) Monte Carlo Tree Search-Based Test Case Mutation.} After coverage growth plateaus, this stage employs MCTS to iteratively select seed queries and apply mutation rules guided by coverage feedback from test execution. With elaborate mutation rules covering schema, data, and query operations, this stage systematically explores deeper execution paths to boost code coverage.

We evaluate MIST on three widely-used open-source DBMSs (DuckDB, PostgreSQL, and SQLite) with four LLMs of different parameter scales (Qwen2.5-7B~\cite{Qwen2.5}, Llama3.1-8B~\cite{llama3.1}, Qwen2.5-14B~\cite{Qwen2.5}, and Qwen2.5-32B~\cite{Qwen2.5}). Experimental results show that MIST improves the baseline approach by 23.9\% $\sim$ 69.17\% on DuckDB, 18.5\% $\sim$ 54.1\% on PostgreSQL, and 18.5\% $\sim$ 92.9\% on SQLite in terms of average code coverage. Moreover, MIST achieves the highest line coverage in the \textit{Optimizer} module, reaching 69.3\% in DuckDB and 63.4\% in PostgreSQL. These results demonstrate that MIST greatly enhances test effectiveness for industrial DBMS testing using only lightweight, locally deployed LLMs. The source code is released at \url{https://github.com/yujiachen99/DBMSTesting}.

\noindent\textbf{Contributions.} The main contributions of our paper are summarized below:

\begin{itemize}[leftmargin=*]
    \item We propose MIST, a novel framework that effectively generates high-quality test cases to improve code coverage for proprietary DBMSs using only lightweight, locally deployed LLMs in resource-constrained industrial environments.
    
    \item We introduce a hierarchical feature tree with error-driven feedback and an MCTS-based mutation engine that jointly optimizes test case generation to improve code coverage.
    
    \item Extensive experiments on three widely-used DBMSs with four LLMs demonstrate that MIST remarkably outperforms baseline models in code coverage and critical module coverage.
    \end{itemize}

{\bf Outline.} 
We present the necessary background in \Cref{sec:background}.
\Cref{sec:overview} presents the overall framework of MIST, followed by detailed introduction. The experimental setup and results are provided in \Cref{sec:exp_setup} and \Cref{sec:result}, respectively.
\Cref{sec:discuss} offers a case study illustrating why MIST works and discusses threats to validity. Finally, we review related work in \Cref{sec:related} and conclude the paper in \Cref{sec:conclusions}.

\section{Background}

\label{sec:background}

\begin{figure*}[t]
    \centering
    \includegraphics[scale=0.48]{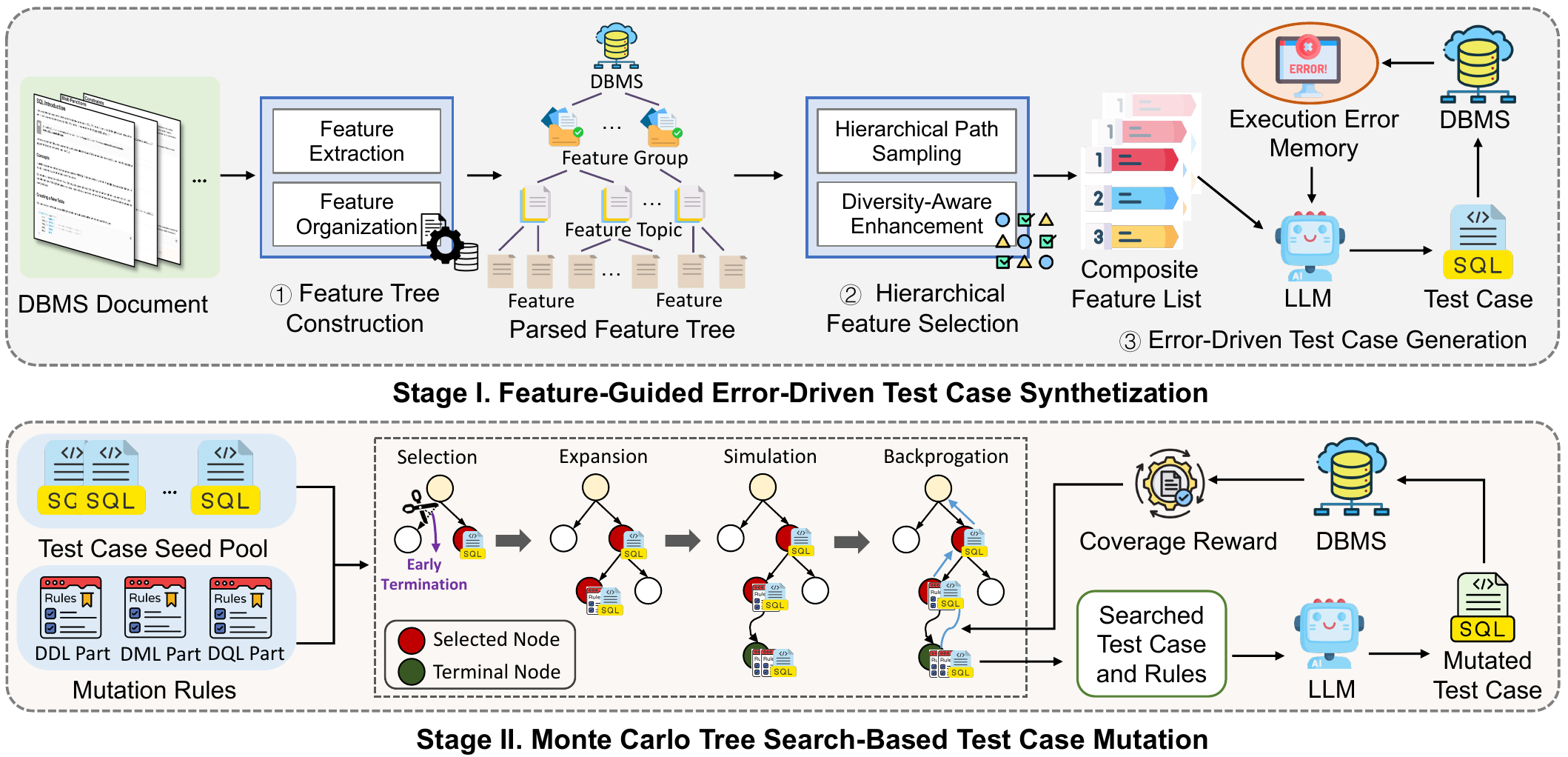}
    \caption{The overview of MIST.}
    \label{fig:framework}
\end{figure*}

\subsection{DBMS Testing}

DBMS testing is a critical process for ensuring the correctness and reliability of database systems, which serve as the backbone of modern data-driven applications. The goal of DBMS testing is to systematically generate SQL test cases that exercise various functional components of the system, including the parser, optimizer, executor, and storage engine, to uncover potential bugs before deployment. 

A typical SQL test case consists of three essential components, as illustrated in Figure~\ref{fig:intro_case}: \textit{Data Definition Language (DDL)} statements that define the database schema (e.g., \texttt{CREATE TABLE} to create tables with specific columns and constraints), \textit{Data Manipulation Language (DML)} statements that populate the database with test data (e.g., \texttt{INSERT} to add records), and \textit{Data Query Language (DQL)} statements that query the database to trigger various execution paths (e.g., \texttt{SELECT} with complex predicates, joins, and aggregations). The interplay of these three components enables comprehensive testing of the DBMS under diverse scenarios. The effectiveness of DBMS testing is measured by \textit{code coverage}, which quantifies how thoroughly the test cases exercise different execution paths in the system. Higher coverage indicates that more code regions have been tested, thereby increasing the likelihood of discovering bugs hidden in less frequently executed paths.

\subsection{Monte Carlo Tree Search}

Monte Carlo Tree Search (MCTS)~\cite{kocsis2006bandit} is a powerful heuristic search algorithm for navigating decision-making problems, particularly those characterized by vast and complex search spaces. Its effectiveness has been demonstrated in a variety of domains~\cite{swiechowski2023monte,zhou2021learned,kartal2019action}. The core of MCTS is an iterative process that asymmetrically grows a search tree by focusing on the most promising regions. Each iteration involves four fundamental steps:

\begin{itemize}
    \item \textbf{Selection:} Starting from the root, the algorithm recursively selects child nodes with the highest utility value until a leaf node is reached. This selection strategy, often guided by the Upper Confidence Bound (UCB) formula, strategically balances the trade-off between exploiting known promising paths and exploring less-visited nodes.
    \item \textbf{Expansion:} Once a leaf node is selected, the tree is expanded by creating one or more child nodes corresponding to new, unexplored actions or states.
    \item \textbf{Simulation:} From a newly expanded node, a lightweight simulation, often called a "rollout," is executed. This typically involves making random or semi-random choices until a terminal state is reached, providing an estimated outcome or value.
    \item \textbf{Backpropagation:} The result from the simulation is then propagated back up the tree, updating the statistics (such as visit counts and value estimates) of all nodes along the path from the expanded node to the root.
\end{itemize}

By repeating this cycle, MCTS intelligently allocates computational resources to the most valuable parts of the search space, making it an ideal technique for optimization problems where exhaustive search is infeasible.

\section{Approach}
\label{sec:overview}
In this section, we propose MIST, an LLM-based test case generation framework for DBMS through Monte Carlo tree search. We first present the overview of MIST and then describe its details in the following subsections.

\subsection{Overview}

Given a target DBMS and a lightweight LLM, MIST generates high-quality SQL test cases through a two-stage approach, as shown in Figure~\ref{fig:framework}. In the (1) \textit{Feature-Guided Error-Driven Test Case Synthetization} stage, MIST constructs a hierarchical feature tree from official documentation and samples feature paths to guide LLM generation. Execution errors are captured and fed back to refine subsequent generations, progressively improving the syntactic validity and semantic diversity of test cases. When coverage growth plateaus, MIST transitions to the (2) \textit{Monte Carlo Tree Search-Based Test Case Mutation} stage, where it treats generated queries as seeds and employs MCTS to iteratively select seeds and apply mutation rules. Coverage feedback from test execution guides the search toward unexplored code regions, systematically improving coverage beyond the initial generation plateau.

\begin{figure}[t]
    \centering
    \includegraphics[scale=1]{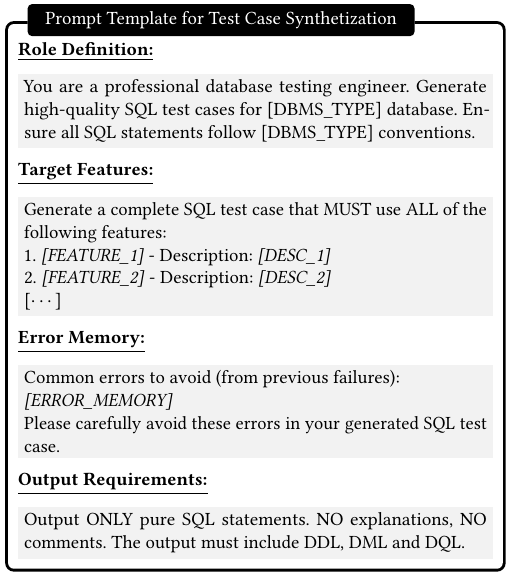}
    \caption{The prompt template for test case synthetization. \textit{[DBMS\_TYPE]} is the target DBMS, \textit{[FEATURE\_LIST]} contains selected features from the hierarchical tree, and \textit{[ERROR\_MEMORY]} provides feedback from previous execution failures.}
    \label{fig:prompt_test}
\end{figure}

\subsection{Feature-Guided Error-Driven Test Case Synthetization}

As shown in Stage I of Figure~\ref{fig:framework}, this stage generates SQL test cases by guiding the LLM with DBMS-specific features. MIST first constructs a hierarchical feature tree from official documentation, then samples feature combinations to construct prompts for generation. Execution error memory enables the LLM to learn dialect-specific constraints and progressively improve the syntactic validity and semantic diversity of generated test cases.

\subsubsection{Feature tree construction}

We collect official documentation from target DBMSs, where each document describes SQL features with syntax specifications and usage examples. These documents cover the full spectrum of SQL dialects, including data types, operators, functions, and query constructs.

\textbf{Feature extraction.} The documentation manager parses documentation files to extract feature entries. Each entry contains a feature name, its functional description, and representative syntax patterns. For instance, DuckDB's string concatenation operator \texttt{||} is extracted with its description and usage pattern \texttt{SELECT 'A' || 'B'}. Similarly, PostgreSQL's \texttt{CONCAT} function is extracted as a distinct feature, capturing the syntactic divergence between DBMSs for the same functionality.

\textbf{Feature organization.} Extracted features are organized into a three-level hierarchical tree, as shown in Figure~\ref{fig:framework}. The first level represents different DBMSs, the second level categorizes features into functional groups (e.g., String Operations, Aggregation Functions), and the third level contains specific feature topics within each group. This parsed feature tree captures the multi-level semantic organization of DBMS functionalities, linking each database with its functional categories and concrete SQL features, thereby enabling fine-grained and context-aware feature sampling.

The feature extraction process is performed manually by parsing official documentation and organizing features into the hierarchical tree structure. For the three evaluated DBMSs, we extract a total of 143 SQL features for DuckDB (covering basic operations, joins/aggregations, advanced features, and specific functionalities), 167 features for PostgreSQL (including basic operations, advanced queries, transactions/views, and specific extensions), and 62 features for SQLite (spanning basic syntax, advanced queries, joins, constraints/indexes, transactions/triggers, and advanced features). Each feature entry contains the feature name, functional description, and representative syntax patterns extracted from the official documentation. The manual construction process for each DBMS takes approximately 6-8 hours, including documentation review, feature extraction, and tree organization. Once constructed, the feature tree can be reused across all experiments and extended incrementally when new DBMS versions introduce additional features.

\subsubsection{Hierarchical feature selection}

To guide the LLM in generating diverse and semantically coherent SQL queries, we employ a two-step feature selection strategy that samples feature combinations from the hierarchical tree.

\textbf{Hierarchical path sampling.} We first randomly select a main path from the feature tree, traversing from a specific DBMS through a feature group to individual features. For example, a sampled path might be ``DuckDB → String Operations → \texttt{||} operator''. We then sample additional features within the same path and expand to related paths within the same DBMS to maintain semantic consistency. This ensures that selected features are functionally related and can be meaningfully combined in a single SQL query.

\textbf{Diversity-aware enhancement.} To increase functional diversity, we introduce cross-category features that are not selected in the first step. For instance, if the main path focuses on string operations, we may add features from aggregation or join operations. This enhancement prevents generated queries from concentrating in a single functional area. The resulting composite feature list balances semantic coherence and functional diversity, providing comprehensive guidance for LLM generation.

\subsubsection{Error-Driven test case generation}

With the selected feature combinations, we construct a structured input prompt to guide the LLM in generating SQL test cases. The prompt template, illustrated in Figure~\ref{fig:prompt_test}, includes four components: (1) a role definition instructing the LLM to act as a database testing engineer, (2) the selected feature list with descriptions, (3) error memory containing common mistakes from previous generations, and (4) output requirements specifying that the response should contain only executable SQL statements including DDL, DML, and DQL. This structured prompt provides both positive guidance through features to use and negative guidance through errors to avoid. The generated test cases are then executed on the target DBMS, and any encountered errors are captured and stored in the memory module. The accumulated error feedback in this memory subsequently guides the LLM to refine and generate more accurate test cases, forming a closed feedback loop that enables MIST to learn from past errors and improve generation quality continuously.

MIST employs an implicit oracle approach where test cases are considered to pass if they execute without crashes, runtime exceptions, or syntax errors. Specifically, we define three types of test failures: (1) \textit{syntax errors} detected during SQL parsing, (2) \textit{runtime exceptions} raised during query execution (e.g., type mismatch, constraint violations), and (3) \textit{system crashes} that terminate the DBMS process. When a test case fails, the error message is captured and stored in the error memory module. This accumulated error feedback guides subsequent LLM generations to avoid similar mistakes, enabling MIST to progressively improve the syntactic validity and semantic correctness of generated test cases. While this implicit oracle approach cannot verify functional correctness (i.e., whether query results are semantically correct), it is highly effective for discovering crashes, parser bugs, and executor errors, which are critical for DBMS reliability.

\subsection{Monte Carlo Tree Search-Based Test Case Mutation}

After the first stage generates an initial set of test cases, coverage growth typically plateaus as new queries become repetitive. To systematically explore deeper execution paths and maximize code coverage, we employ Monte Carlo Tree Search (MCTS) to guide the mutation process, as shown in Stage II of Figure~\ref{fig:framework}.

\subsubsection{Mutation rules curation}

We formulate test case mutation as a tree-structured search problem over a directed tree $\mathcal{T} = (V, E)$. In this formulation, each node represents a test case, while each edge corresponds to an action, either selecting a seed SQL from the test case pool or applying a specific mutation rule. Therefore, each root-to-leaf path represents a mutation trajectory, i.e., a complete sequence of selections and transformations that progressively evolve an initial seed test case into a newly mutated one. By employing MCTS, we can jointly optimize the selection of seed test cases and the application of mutation rules, thereby discovering the most coverage-effective mutations.


To support this search process, we curate a collection of 135 dialect-aware mutation rules (45 rules per DBMS) covering schema, data, and query operations. For each DBMS, the rules are distributed as follows: 10 DDL rules that modify database schemas (e.g., adding columns with complex data types, creating indexes), 10 DML rules that manipulate data (e.g., inserting boundary values, NULL handling), and 25 DQL rules that transform queries (e.g., changing join types, adding subqueries, introducing window functions). Together, these rule categories define a comprehensive mutation space that allows MIST to systematically transform SQL test cases while preserving syntactic validity.



\begin{algorithm}[t]
\caption{{\tt MCTS for test case mutation}}
\label{alg:traditional-mcts}
\KwIn{The number of iterations $T$}
\KwOut{Mutated test case with maximum reward}
Initialize MCTS tree root $v_0$\;
\ForEach{$t \gets 1, 2, 3, \cdots, T$}{
  $v_{t} \gets \texttt{Select}(v_0)$ \;
  
  $(v_{t}', a_t) \gets \texttt{Expand}(v_{t})$ \;

  \tcp{\textcolor{teal}{Complete the mutation trajectory randomly}}
  $\theta_{t} \gets \texttt{Rollout}(v_{t}')$ \;
  
  $ \widehat{R}(v_t, a_t) \gets f_{Q}(\theta_{t})$ \;
  {Backpropagation via equations (2) - (4)}\;

}
\end{algorithm}






\subsubsection{MCTS-based mutation process} We employ MCTS to iteratively explore the mutation search space and discover high-coverage test cases. As detailed in Algorithm~\ref{alg:traditional-mcts}, MCTS operates through four fundamental steps that work together to balance exploration of new mutations and exploitation of promising trajectories. At each iteration, the algorithm starts from the root node and progressively builds the search tree by selecting actions, expanding nodes, simulating complete mutation trajectories, and updating node statistics based on coverage feedback from the target DBMS.

\textbf{Selection.} At each iteration, MCTS traverses the search tree from the root $v_0$ to a leaf $v_t$ by recursively selecting the child node with the highest score (line~3 in Algorithm~\ref{alg:traditional-mcts}), where the Upper Confidence Bound for Trees (UCT)~\cite{kocsis2006bandit} is used as the scoring policy.

Specifically, for a state $v$ and an action $a \in \mathcal{A}(v)$, where $\mathcal{A}(v)$ denotes the set of feasible actions from $v$, the UCT score is:
\begin{equation}
    U(v,a) = \frac{R(v,a)}{N(v,a)} \;+\; c \cdot \sqrt{\frac{\ln N(v)}{N(v,a)}},
\end{equation}
where $R(v,a)$ is the cumulative reward obtained by executing action $a$ at state $v$, $N(v,a)$ is the number of times action $a$ has been taken from $v$, and $N(v)$ is the total number of visits to $v$.  The constant $c$ balances exploration and exploitation.

\textbf{Expansion.}
In the $t$-th iteration, if the selected node $v_t$ is not at the terminal layer or is not fully expanded (i.e., there exists an action in $\mathcal{A}(v_t)$ that has not yet been executed), we randomly choose one such unexplored action, denoted as $a_t$. Executing $a_t$ generates a new child state $v_t'$, which is then added to the search tree (line 4).

\begin{figure}[t]
    \centering
    \includegraphics[scale=0.6]{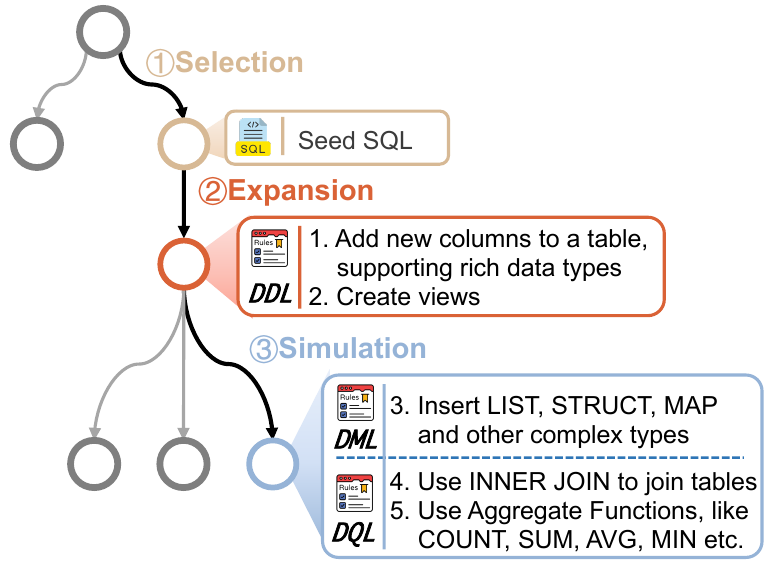}
    \caption{Illustrating the MCTS for test case mutation.}
    \label{fig:MCTS_case_study}
\end{figure}

\textbf{Simulation.} In the simulation phase (lines 5–6), suppose the newly expanded node $v_t'$ lies at layer $i$. The algorithm then performs a rollout strategy to complete the remaining mutation trajectory from $v_t'$ to a terminal node. A commonly used strategy is random playout, in which the remaining mutation actions are sampled uniformly at random and combined with the already fixed actions along the path leading to $v_t'$, thereby producing a complete mutated test case for evaluation.

\textbf{Backpropagation.} Following the simulation phase, the final reward $\widehat{R}(v_t, a_t) = f(\theta_t)$ is propagated back along the search path, where $f(\theta_t)$ denotes the coverage rate achieved by executing the generated test cases $\theta_t$ on the current DBMS under test. The goal of this backpropagation phase is to update the value estimates and visit counts for all intermediate state–action pairs encountered during the rollout, thereby refining the search policy toward actions that yield higher coverage in subsequent iterations. Specifically, for each state-action pair $(v, a)$ along the search path, we update three statistics:
\begin{align}
    R(v, a) & \leftarrow R(v, a) + \widehat{R}(v, a) \\
    N(v)      & \leftarrow N(v) + 1 \\
    N(v, a) & \leftarrow N(v, a) + 1
\end{align}
where $R(v,a)$ is the cumulative reward, $N(v)$ is the visit count of node $v$, and $N(v,a)$ is the visit count of action $a$ from node $v$.

\begin{example}
Figure~\ref{fig:MCTS_case_study} illustrates an example of MCTS for test case mutation. Each iteration begins at the root node. The algorithm traverses the tree by repeatedly selecting the node with the highest UCT score (i.e., selecting a seed SQL) until it reaches an expandable node. From there, the algorithm expands the tree by randomly choosing an untried action, i.e., selecting a DDL mutation rule, to create a new node.
Next, a simulation (or rollout) is performed from this state, randomly applying additional actions from the DML and DQL rule sets to complete the mutation trajectory. The resulting mutated SQL query is then executed on the target DBMS to obtain a reward (i.e., the achieved coverage).
Finally, this reward is used to update the statistics of all nodes along the path back to the root, which is highlighted the black in this figure.
\end{example}

\textbf{Early termination.} To improve search efficiency, we design an early termination strategy that prioritizes seed SQLs with recently high performance while automatically pruning saturated or over-complex ones. Specifically, the strategy continuously tracks the recent branch coverage improvement of each SQL; once its coverage gain becomes saturated, the corresponding node is excluded from further expansion in subsequent MCTS iterations.

\section{Experimental Setup}
\label{sec:exp_setup}
\subsection{Research Questions}

In this paper, we mainly investigate the following research questions through experiments.

\begin{itemize}

    \item \textbf{RQ1}: How effective is MIST in improving code coverage for DBMSs compared to baseline approaches?

    \item \textbf{RQ2}: How does MIST perform across different DBMS modules (i.e., \textit{Parser}, \textit{Optimizer}, \textit{Executor}, \textit{Storage})?
    
    \item \textbf{RQ3}: What are the impacts of the two stages in MIST on code coverage improvement?
       
\end{itemize}

\subsection{Studied DBMSs}

Following prior studies~\cite{zhong2025scaling,zhong2025testing}, we evaluate MIST on three widely-used open-source DBMS representing diverse architectural designs and SQL dialect characteristics.

\begin{itemize}

\item \textbf{DuckDB~\cite{raasveldt2019duckdb}:} An in-process analytical database designed for OLAP workloads with columnar storage and vectorized execution. It supports modern SQL features including window functions and CTEs.

\item \textbf{PostgreSQL~\cite{stonebraker1986design}:} A mature relational database system with a sophisticated query optimizer and advanced features such as materialized views and JSON operations.

\item \textbf{SQLite~\cite{owens2010sqlite}:} A lightweight embedded database engine widely deployed in mobile and embedded systems. It implements a comprehensive SQL dialect with unique features such as virtual tables.

\end{itemize}

These three systems collectively cover a broad spectrum of DBMS architectures (in-process analytical, client-server relational, and embedded transactional, respectively), enabling us to comprehensively evaluate MIST's generalizability across different SQL dialects and execution models.



\subsection{Studied LLMs and Baseline}

To evaluate MIST's effectiveness across different model capacities, we select four open-source LLMs spanning different parameter scales and model families: Qwen2.5-7B, Llama3.1-8B, Qwen2.5-14B, and Qwen2.5-32B. These models represent lightweight to medium-sized LLMs that are practical for local deployment in industrial environments, aligning with the resource constraints commonly faced in real-world DBMS testing scenarios.

For baseline comparison, we select Fuzz4All~\cite{fuzz4all} as our primary baseline. To the best of our knowledge, there is currently no existing work that uses LLMs as complete test case generators for DBMS testing. Inspired by ShQveL~\cite{zhong2025testing}, we adopt the state-of-the-art LLM-based framework Fuzz4All for fair comparison. Fuzz4All is the first universal fuzzer that leverages LLMs to generate test inputs across multiple programming languages and systems.  

\subsection{Metrics}

We evaluate MIST using two categories of coverage metrics to comprehensively assess its effectiveness in exercising DBMS code.

\begin{table*}[]
\caption{Overall code coverage comparison between baseline and MIST.}
\label{table:rq1}
\begin{tabular}{c|c|ccc|ccc|ccc}
\toprule
 &  & \multicolumn{3}{c|}{\textbf{DuckDB}} & \multicolumn{3}{c|}{\textbf{PostgreSQL}} & \multicolumn{3}{c}{\textbf{SQLite}} \\ \cline{3-11} 
\multirow{-2}{*}{\textbf{Model}} & \multirow{-2}{*}{\textbf{Approach}} & \textbf{Line} & \textbf{Function} & \textbf{Branch} & \textbf{Line} & \textbf{Function} & \textbf{Branch} & \textbf{Line} & \textbf{Function} & \textbf{Branch} \\ \midrule
 & Fuzz4All & 29.9\% & 29.7\%  & 17.0\% & 23.1\% & 25.2\% & 16.2\% & 29.9\% & 38.9\%  & 22.5\% \\
\multirow{-2}{*}{Qwen2.5-7B} & \cellcolor[HTML]{ECF4FF}\textbf{MIST} & \cellcolor[HTML]{ECF4FF}\textbf{42.0\%} & \cellcolor[HTML]{ECF4FF}\textbf{38.2\%} & \cellcolor[HTML]{ECF4FF}\textbf{24.5\%} & \cellcolor[HTML]{ECF4FF}\textbf{30.3\%} & \cellcolor[HTML]{ECF4FF}\textbf{38.6\%} & \cellcolor[HTML]{ECF4FF}\textbf{21.3\%} & \cellcolor[HTML]{ECF4FF}\textbf{39.7\%} & \cellcolor[HTML]{ECF4FF}\textbf{46.1\%} & \cellcolor[HTML]{ECF4FF}\textbf{31.3\%} \\ \midrule
 & Fuzz4All & 32.3\% & 31.7\% & 18.3\% & 24.2\% & 31.8\% & 16.9\% & 28.3\% & 36.6\% & 21.1\% \\
\multirow{-2}{*}{Llama3.1-8B} & \cellcolor[HTML]{ECF4FF}\textbf{MIST} & \cellcolor[HTML]{ECF4FF}\textbf{43.2\%} & \cellcolor[HTML]{ECF4FF}\textbf{39.3\%} & \cellcolor[HTML]{ECF4FF}\textbf{25.4\%} & \cellcolor[HTML]{ECF4FF}\textbf{29.3\%} & \cellcolor[HTML]{ECF4FF}\textbf{37.7\%} & \cellcolor[HTML]{ECF4FF}\textbf{20.5\%} & \cellcolor[HTML]{ECF4FF}\textbf{39.9\%} & \cellcolor[HTML]{ECF4FF}\textbf{45.2\%} & \cellcolor[HTML]{ECF4FF}\textbf{32.1\%} \\ \midrule
 & Fuzz4All & 26.4\% & 26.8\% & 14.6\% & 23.6\% & 31.3\% & 16.4\% & 21.6\% & 30.3\% & 15.6\% \\
\multirow{-2}{*}{Qwen2.5-14B} & \cellcolor[HTML]{ECF4FF}\textbf{MIST} & \cellcolor[HTML]{ECF4FF}\textbf{42.2\%} & \cellcolor[HTML]{ECF4FF}\textbf{38.6\%} & \cellcolor[HTML]{ECF4FF}\textbf{24.7\%} & \cellcolor[HTML]{ECF4FF}\textbf{31.4\%} & \cellcolor[HTML]{ECF4FF}\textbf{39.9\%} & \cellcolor[HTML]{ECF4FF}\textbf{22.2\%} & \cellcolor[HTML]{ECF4FF}\textbf{38.1\%} & \cellcolor[HTML]{ECF4FF}\textbf{44.5\%} & \cellcolor[HTML]{ECF4FF}\textbf{30.1\%} \\ \midrule
 & Fuzz4All & 25.8\% & 28.0\% & 18.6\% & 20.9\% & 28.4\% & 14.4\% &26.5\%  &34.7\%  & 19.8\% \\
\multirow{-2}{*}{Qwen2.5-32B} & \cellcolor[HTML]{ECF4FF}\textbf{MIST} & \cellcolor[HTML]{ECF4FF}\textbf{41.8\%} & \cellcolor[HTML]{ECF4FF}\textbf{38.6\%} & \cellcolor[HTML]{ECF4FF}\textbf{24.4\%} & \cellcolor[HTML]{ECF4FF}\textbf{31.5\%} & \cellcolor[HTML]{ECF4FF}\textbf{40.1\%} & \cellcolor[HTML]{ECF4FF}\textbf{22.2\%} & \cellcolor[HTML]{ECF4FF}\textbf{36.5\%} & \cellcolor[HTML]{ECF4FF}\textbf{43.1\%} & \cellcolor[HTML]{ECF4FF}\textbf{29.0\%} \\ \bottomrule
\end{tabular}
\end{table*}

\begin{itemize}

    \item \textbf{Code Coverage.} We measure three standard code coverage metrics: (1) \textit{Line Coverage}, which calculates the percentage of executed source code lines; (2) \textit{Function Coverage}, which measures the percentage of functions that are invoked at least once; and (3) \textit{Branch Coverage}, which tracks the percentage of conditional branches (e.g., if-else statements) that are executed. These metrics provide a holistic view of how thoroughly the generated test cases exercise the DBMS codebase.

    \item \textbf{Module-Level Coverage.} To understand the fine-grained impact of MIST on different DBMS components, we also measure coverage across four critical functional modules: (1) \textit{Parser}, which handles SQL parsing and syntax validation; (2) \textit{Optimizer}, which performs query optimization and plan generation; (3) \textit{Executor}, which executes the query plan and produces results; and (4) \textit{Storage}, which manages data storage and retrieval operations. This module-level analysis reveals which components benefit most from MIST's test generation strategy.

\end{itemize}

\subsection{Implementation Details}

All LLMs are downloaded from Hugging Face~\cite{Huggingface} and deployed locally using vLLM~\cite{kwon2023efficient} as the inference engine on 8 NVIDIA GPUs, with generation parameters configured as follows: maximum token length of 1,024, temperature of 0.2 for stable generation, and batch size of 1 with 3 concurrent workers to balance throughput and resource utilization. In Stage I, we generate 900 initial SQL test cases for each DBMS, with each test case incorporating 3 to 20 selected features from the hierarchical feature tree to ensure diversity. The range of 3 to 20 features is chosen through preliminary experiments to balance semantic richness and syntactic complexity: fewer than 3 features tend to produce overly simple queries with limited coverage, while more than 20 features often result in syntactically invalid SQL due to feature conflicts. In Stage II, we perform 600 mutation iterations to explore deeper execution paths and improve coverage incrementally. After generation, we apply post-processing to ensure the generated SQL can be correctly executed, including removing Markdown code block markers, stripping explanatory text and comments, extracting only valid SQL statements (DDL, DML, and DQL), and ensuring proper semicolon termination following standard practices in LLM-based code generation~\cite{joel2024survey,wang2023review,liu2024exploring}. For the MCTS mutation, we set the exploration weight $c$ in the UCT formula to 1.414 following the standard UCT formula recommendation~\cite{kocsis2006bandit}, and define an early-termination threshold of 50 new branches, which is empirically set to balance exploration efficiency and coverage improvement potential.

\section{Result}

\label{sec:result}
\subsection{Code Coverage of MIST (RQ1)}

\textbf{Experimental Design.} To answer this research question, we evaluate MIST on three DBMSs (DuckDB, PostgreSQL, SQLite) using four LLMs (Qwen2.5-7B, Llama3.1-8B, Qwen2.5-14B, Qwen2.5-32B).

\textbf{Results Analysis.} Table~\ref{table:rq1} shows the coverage comparison between the baseline and MIST. Based on the results, we make the following observations.

\textbf{(1) MIST consistently improves coverage across all DBMSs and coverage metrics.} Across the three DBMSs, MIST achieves substantial coverage improvements over the baseline. For DuckDB, MIST improves line coverage by an average of 49.0\%, function coverage by 33.6\%, and branch coverage by 45.8\%. For PostgreSQL, the improvements are 34.0\% in line coverage, 35.1\% in function coverage, and 35.6\% in branch coverage. For SQLite, MIST achieves improvements of 47.0\% in line coverage, 28.3\% in function coverage, and 57.7\% in branch coverage. These consistent improvements across diverse DBMS architectures, ranging from in-process analytical (DuckDB) to client-server relational (PostgreSQL) and embedded transactional (SQLite), demonstrate that MIST is effective regardless of the underlying database design.

\textbf{(2) MIST shows strong generalization across different LLM sizes and families.} The effectiveness of MIST is consistent across all four evaluated LLMs, regardless of parameter scale or model family. For the Qwen family, MIST achieves branch coverage improvements ranging from 31.2\% to 69.2\% on DuckDB, 31.5\% to 54.2\% on PostgreSQL, and 39.1\% to 92.9\% on SQLite. For Llama3.1-8B, MIST achieves improvements of 38.8\%, 21.3\%, and 52.1\% on the three DBMSs, respectively. Notably, smaller models with MIST often outperform larger baseline models. For example, Qwen2.5-7B with MIST achieves 42.0\% line coverage on DuckDB, surpassing the baseline of Qwen2.5-32B (25.8\%). This demonstrates that MIST is more critical than raw model capacity, making it practical for resource-constrained industrial environments where deploying large models may not be feasible.


\noindent
\fcolorbox{gray!15}{gray!15}{
\parbox{\dimexpr\linewidth-2\fboxsep-2\fboxrule}{
\textbf{Answer to RQ1:} MIST consistently improves code coverage across all three DBMSs and four LLMs, achieving average improvements of 43.3\% in line coverage, 32.3\% in function coverage, and 46.4\% in branch coverage compared to the baseline approach. The improvements demonstrate strong generalization across different DBMS architectures and LLM sizes.
}}

\begin{table}[t]
\caption{Module-level line coverage comparison between baseline and MIST.}
\label{table:rq2}
\small
\centering
\begin{tabular}{c|c|cccc}
\toprule
\textbf{Model} & \textbf{Approach} & \textbf{Parser} & \textbf{Optimizer} & \textbf{Executor} & \textbf{Storage} \\ 
\midrule

\multicolumn{6}{c}{\cellcolor[HTML]{ECF4FF}\textbf{DuckDB}} \\ \midrule

\multirow{2}{*}{\begin{tabular}[c]{@{}c@{}}Qwen2.5\\ 7B\end{tabular}}
  & Fuzz4All & 26.6\% & 51.6\% & 31.0\% & 20.0\% \\
  & \textbf{MIST} & \textbf{43.2\%} & \textbf{68.4\%} & \textbf{42.2\%} & \textbf{31.3\%} \\ \midrule

\multirow{2}{*}{\begin{tabular}[c]{@{}c@{}}Llama3.1\\ 8B\end{tabular}}
  & Fuzz4All  & 30.1\% & 52.8\% & 35.1\% & 22.7\% \\
  & \textbf{MIST} & \textbf{48.9\%} & \textbf{69.3\%} & \textbf{47.3\%} & \textbf{32.9\%} \\ \midrule

\multirow{2}{*}{\begin{tabular}[c]{@{}c@{}}Qwen2.5\\ 14B\end{tabular}}
  & Fuzz4All  & 24.2\% & 43.4\% & 22.9\% & 22.1\% \\
  & \textbf{MIST} & \textbf{45.1\%} & \textbf{67.0\%} & \textbf{46.6\%} & \textbf{31.8\%} \\ \midrule

\multirow{2}{*}{\begin{tabular}[c]{@{}c@{}}Qwen2.5\\ 32B\end{tabular}}
  & Fuzz4All  & 23.7\% & 45.6\% & 26.4\% & 22.6\% \\
  & \textbf{MIST} & \textbf{43.0\%} & \textbf{66.9\%} & \textbf{45.2\%} & \textbf{31.8\%} \\ \midrule

\multicolumn{6}{c}{\cellcolor[HTML]{ECF4FF}\textbf{PostgreSQL}} \\ \midrule

\multirow{2}{*}{\begin{tabular}[c]{@{}c@{}}Qwen2.5\\ 7B\end{tabular}}
  & Fuzz4All  & 36.1\% & 50.8\% & 37.5\% & 31.8\% \\
  & \textbf{MIST} & \textbf{41.0\%} & \textbf{59.0\%} & \textbf{43.1\%} & \textbf{36.4\%} \\ \midrule

\multirow{2}{*}{\begin{tabular}[c]{@{}c@{}}Llama3.1\\ 8B\end{tabular}}
  & Fuzz4All  & 32.9\% & 47.3\% & 33.5\% &  30.9\% \\
  & \textbf{MIST} & \textbf{39.9\%} & \textbf{57.4\%} & \textbf{39.1\%} & \textbf{36.3\%} \\ \midrule

\multirow{2}{*}{\begin{tabular}[c]{@{}c@{}}Qwen2.5\\ 14B\end{tabular}} 
  & Fuzz4All  & 32.2\% & 44.6\%  & 30.6\% &  30.5\% \\
  & \textbf{MIST} & \textbf{41.5\%} & \textbf{62.6\%} & \textbf{45.8\%} & \textbf{38.1\%} \\ \midrule

\multirow{2}{*}{\begin{tabular}[c]{@{}c@{}}Qwen2.5\\ 32B\end{tabular}}
  & Fuzz4All  & 29.6\% & 38.4\% & 25.3\% & 27.9\% \\
  & \textbf{MIST} & \textbf{41.2\%} & \textbf{63.4\%} & \textbf{45.9\%} & \textbf{38.1\%} \\ 

\bottomrule
\end{tabular}
\end{table}

\begin{table*}[t]
\caption{Ablation study results showing the impact of different components in MIST. Each row represents a different combination of synthetization strategy and mutation strategy.}
\label{table:rq3}
\begin{tabular}{c|c|ccc|ccc|ccc}
\toprule
\multirow{2}{*}{\textbf{Synthetization}} & \multirow{2}{*}{\textbf{Mutation}} & \multicolumn{3}{c|}{\textbf{DuckDB}} & \multicolumn{3}{c|}{\textbf{PostgreSQL}} & \multicolumn{3}{c}{\textbf{SQLite}} \\ \cline{3-11} 
 &  & \textbf{Line} & \textbf{Function} & \textbf{Branch} & \textbf{Line} & \textbf{Function} & \textbf{Branch} & \textbf{Line} & \textbf{Function} & \textbf{Branch} \\ \midrule
\begin{tabular}[c]{@{}c@{}}Simple \\ Instruction\end{tabular} & \XSolidBrush  & 29.9\% & 29.7\%  & 17.0\% & 23.1\% & 25.2\% & 16.2\% & 29.9\% & 38.9\%  & 22.5\%  \\ \midrule
\begin{tabular}[c]{@{}c@{}}Random\\ Feature\end{tabular} & \XSolidBrush & 32.3\% & 31.7\% & 18.7\% & 25.2\% & 30.1\% & 16.9\% & 32.8\% & 41.7\% & 24.6\% \\ \midrule
\begin{tabular}[c]{@{}c@{}}Hierarchical\\ Feature \\ (MIST)\end{tabular} & \XSolidBrush & 36.2\% & 34.7\% & 20.6\% & 27.4\% & 32.1\% & 17.8\% & 35.5\% & 43.2\% & 26.3\%
 \\ \midrule \midrule
\begin{tabular}[c]{@{}c@{}}Hierarchical\\ Feature \\ (MIST)\end{tabular} & \begin{tabular}[c]{@{}c@{}}Simple \\ Instruction\end{tabular} & 37.8\% & 35.7\% & 21.6\% & 28.3\% & 33.9\% & 18.1\%  & 36.9\% & 44.2\% & 27.5\% \\ \midrule
\begin{tabular}[c]{@{}c@{}}Hierarchical\\ Feature \\ (MIST)\end{tabular} & \begin{tabular}[c]{@{}c@{}}Random\\ Rule\end{tabular} & 38.7\% & 36.7\% & 22.4\% & 28.8\% & 35.4\% & 19.5\% & 37.5\% & 44.8\% & 29.2\% \\ \midrule
\textbf{\begin{tabular}[c]{@{}c@{}}Hierarchical\\ Feature \\ (MIST)\end{tabular}} & \textbf{\begin{tabular}[c]{@{}c@{}}MCTS\\ Rule \\ (MIST)\end{tabular}} & \textbf{42.0\%} & \textbf{38.2\%} & \textbf{24.5\%} & \textbf{30.3\%} & \textbf{38.6\%} & \textbf{21.3\%} & \textbf{39.7\%} & \textbf{46.1\%} & \textbf{31.3\%} \\ 

\bottomrule
\end{tabular}
\end{table*}

\subsection{Module-Level Coverage of MIST (RQ2)}
\noindent \textbf{Experimental Design.} To answer this research question, we analyze the line coverage distribution across four critical DBMS modules: \textit{Parser}, \textit{Optimizer}, \textit{Executor}, and \textit{Storage}. We focus on DuckDB and PostgreSQL, as SQLite's monolithic architecture does not provide clear module boundaries for fine-grained analysis.

\noindent \textbf{Results Analysis.} Table~\ref{table:rq2} shows the module-level line coverage comparison between the baseline and MIST. Based on the results, we make the following observations.

\textbf{(1) MIST achieves substantial improvements across all four modules.} For DuckDB, MIST improves \textit{Parser} coverage by an average of 73.2\%, \textit{Optimizer} coverage by 41.3\%, \textit{Executor} coverage by 61.4\%, and \textit{Storage} coverage by 46.5\% across the four LLMs. For PostgreSQL, the improvements are 25.8\% in \textit{Parser}, 35.8\% in \textit{Optimizer}, 40.7\% in \textit{Executor}, and 23.4\% in \textit{Storage}. These comprehensive improvements demonstrate that MIST effectively exercises execution paths across the entire DBMS pipeline, from front-end parsing to back-end storage operations. Notably, even the \textit{Storage} module, which handles low-level data access and buffer management, shows consistent gains of 46.5\% and 23.4\% for DuckDB and PostgreSQL, respectively, indicating that MIST's test cases successfully trigger diverse storage access patterns.

\textbf{(2) The Optimizer module achieves the highest absolute coverage across both DBMSs.} Among the four modules, the \textit{Optimizer} consistently achieves the highest absolute coverage values with MIST. In DuckDB, MIST reaches an average of 67.9\% \textit{Optimizer} coverage, compared to 48.4\% in the baseline. In PostgreSQL, MIST achieves 60.6\% \textit{Optimizer} coverage, up from 45.3\% in the baseline. This is particularly because the \textit{Optimizer} is one of the most complex components in modern DBMSs, responsible for query plan generation, cost estimation, and selecting optimal execution strategies. The high coverage indicates that MIST's feature-guided generation and MCTS-based mutation effectively trigger diverse optimization logic. For example, Llama3.1-8B with MIST achieves 69.3\% \textit{Optimizer} coverage on DuckDB, demonstrating that even smaller models can thoroughly exercise complex optimization paths when guided by MIST.

\textbf{(3) Different modules benefit differently from MIST, reflecting their architectural characteristics.} The magnitude of improvement varies across modules, with \textit{Parser} and \textit{Executor} showing larger relative gains in DuckDB (73.2\% and 61.4\%), while PostgreSQL shows more balanced improvements across all modules (23.4\% to 40.7\%). This difference reflects DuckDB's specialized design for analytical workloads, where \textit{Parser} and \textit{Executor} handle complex operations like window functions, CTEs, and vectorized execution. In contrast, PostgreSQL's mature and extensively tested codebase shows more uniform improvements, indicating that MIST discovers previously untested paths across all components rather than concentrating on specific modules.


\noindent
\fcolorbox{gray!15}{gray!15}{
\parbox{\dimexpr\linewidth-2\fboxsep-2\fboxrule}{
\textbf{Answer to RQ2:} MIST achieves substantial improvements across all DBMS modules, with the \textit{Optimizer} module showing the highest absolute coverage (average 67.9\% for DuckDB and 60.6\% for PostgreSQL).
}}

\subsection{Ablation Study (RQ3)}

\noindent \textbf{Experimental Design.} To answer this research question, we perform ablation studies by evaluating different combinations of synthetization and mutation strategies using Qwen2.5-7B. For synthetization, we compare three approaches:

\begin{itemize}
\item \textbf{Simple Instruction:} Uses basic prompts without feature guidance.
\item \textbf{Random Feature:} Randomly selects features without hierarchical structure.
\item \textbf{Hierarchical Feature:} Uses our hierarchical feature tree with diversity-aware sampling.
\end{itemize}

For mutation, we compare four approaches:

\begin{itemize}
\item \textbf{No Mutation:} Skips the mutation stage.
\item \textbf{Simple Instruction:} Uses basic prompts for mutation.
\item \textbf{Random Rule:} Randomly applies mutation rules.
\item \textbf{MCTS Rule:} Uses our MCTS-based mutation engine.
\end{itemize}

\noindent \textbf{Results Analysis.} Table~\ref{table:rq3} shows the coverage results of different component combinations. Based on the results, we make the following observations.

\textbf{(1) Hierarchical feature selection is the most effective synthetization strategy.} Comparing the three synthetization strategies without mutation, \textit{Hierarchical Feature} achieves branch coverage of 20.6\% on DuckDB, 17.8\% on PostgreSQL, and 26.3\% on SQLite. This outperforms \textit{Random Feature} by 10.2\%, 5.3\%, and 6.9\%, respectively, and \textit{Simple Instruction} by 21.2\%, 9.9\%, and 16.9\%, respectively. The superior performance of \textit{Hierarchical Feature} can be attributed to its ability to systematically sample features from the hierarchical tree while maintaining semantic coherence, ensuring that generated test cases exercise diverse DBMS functionalities. 

\begin{figure*}[t]
    \centering
    \includegraphics[scale=0.55]{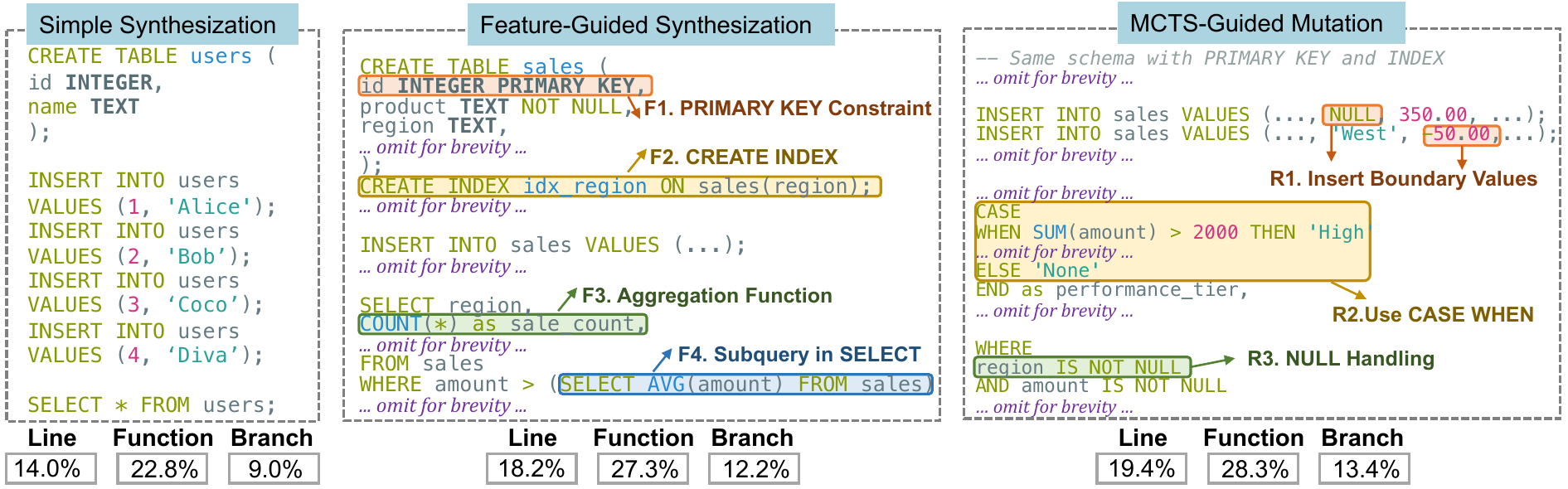}
    \caption{Case study on test case generation using Qwen2.5-7B on SQLite.}
    \label{fig:case_study}
\end{figure*}

\textbf{(2) MCTS-based mutation consistently outperforms alternative mutation strategies.} Among the three mutation approaches applied to \textit{Hierarchical Feature}, \textit{MCTS Rule} achieves branch coverage of 24.5\% on DuckDB, 21.3\% on PostgreSQL, and 31.3\% on SQLite. This represents improvements of 13.4\%, 17.7\%, and 13.8\% over \textit{Simple Instruction}, and 9.4\%, 9.2\%, and 7.2\% over \textit{Random Rule}, respectively. The consistent superiority of \textit{MCTS Rule} demonstrates that its tree search mechanism effectively balances exploration of new mutation paths and exploitation of high-coverage seeds. \textit{Random Rule} performs better than \textit{Simple Instruction} but still falls short of \textit{MCTS Rule}, indicating that random selection cannot match the systematic search strategy provided by MCTS.

\textbf{(3) Removing either component leads to performance degradation, among which hierarchical feature selection has the greatest impact.} The complete MIST framework (Hierarchical Feature and MCTS Rule) achieves the highest coverage across all DBMSs. Removing the mutation stage (Hierarchical Feature and No Mutation) reduces branch coverage by 18.9\% on DuckDB, 19.7\% on PostgreSQL, and 19.0\% on SQLite. Replacing \textit{Hierarchical Feature} with \textit{Simple Instruction} while keeping \textit{MCTS Rule} would reduce coverage by 44.1\%, 31.5\%, and 39.1\% on the three DBMSs, respectively. These results demonstrate that both components are essential, with hierarchical feature selection being the more critical component.


\noindent
\fcolorbox{gray!15}{gray!15}{
\parbox{\dimexpr\linewidth-2\fboxsep-2\fboxrule}{
\textbf{Answer to RQ3:} Hierarchical feature selection outperforms baselines by 5.3\%-21.2\%, while MCTS-based mutation further improves coverage by 9.2\%-17.7\% over alternative mutation strategies. Removing either component leads to substantial performance degradation, with hierarchical feature selection having the greater impact, confirming the necessity of the complete framework.
}}

\section{Discussion}
\label{sec:discuss}

\subsection{Why does MIST work?}

The effectiveness of MIST lies in guiding LLM generation with hierarchical documentation structure and systematically exploring execution paths through MCTS-based mutation. As demonstrated in Figure~\ref{fig:case_study}, we present a case study using Qwen2.5-7B on SQLite, comparing three approaches to SQL test case generation: simple synthesization, feature-guided synthesization, and MCTS-guided mutation.

\subsubsection{Feature-guided synthetization for enriching the semantics of test cases with document knowledge}

As shown in Figure~\ref{fig:case_study}, \textit{Simple Synthesization} fails to provide sufficient semantic richness. It relies solely on basic table structures and straightforward \texttt{INSERT/SELECT} statements, generating the minimal SQL that achieves only 14.0\% line coverage and 9.0\% branch coverage. This represents the lower bound of what random generation can produce. In contrast, by organizing DBMS features into a hierarchical tree and performing path-based sampling, MIST systematically selects features that exercise deeper components. The feature-guided approach incorporates PRIMARY KEY constraints (F1), CREATE INDEX (F2), aggregation functions (F3), and subqueries (F4) into the generated SQL, resulting in improvements of 30.0\%, 19.7\%, and 35.6\% in terms of line coverage, function coverage, and branch coverage, respectively. This demonstrates that hierarchical feature selection effectively guides LLMs to generate semantically rich SQL that explores wider execution paths.

\subsubsection{Monte Carlo tree search-based mutation for exploring deeper execution paths to enhance coverage}

MIST strategically mutates existing high-quality seeds using rules selected through MCTS, which ensures that mutation rules are systematically chosen to maximize coverage improvement potential rather than being randomly applied. Specifically, the mutation introduces boundary value insertion (R1) with extreme values (NULL, -50.00), CASE WHEN branches (R2) with conditional logic, and NULL handling (R3) with IS NOT NULL predicates. These mutations trigger edge case handling in arithmetic operations and storage management, activate different evaluation paths in the executor, and exercise NULL value processing logic. The MCTS-guided mutation achieves 19.4\% line coverage, 28.3\% function coverage, and 13.4\% branch coverage, representing additional improvements of 6.6\%, 3.7\%, and 9.8\% over feature-guided synthesis, respectively.

The comparison across three approaches demonstrates that feature-guided synthesis and MCTS-based mutation address different aspects of coverage improvement. Feature selection provides semantic richness by guiding LLMs based on documentation knowledge, while MCTS-based mutation provides systematic edge case exploration when generation saturates, demonstrating the effectiveness of MIST in comprehensive DBMS testing.

\subsection{Threats to Validity}

We identify three main threats to the validity of our study:

\noindent\textbf{Internal Validity.} The primary threat concerns coverage measurement accuracy. We use compiler-based instrumentation (gcov/lcov) to collect coverage data, which may introduce runtime overhead. To mitigate this, we use dedicated coverage-enabled builds separate from production binaries and verify that instrumentation does not alter DBMS functional correctness. Another concern is the non-deterministic nature of LLM-based generation. To mitigate this variability, we configure a fixed temperature of 0.2 and use consistent random seeds across all experiments for reproducibility. The substantial and consistent coverage improvements observed across all four LLMs and three DBMSs demonstrate that MIST's effectiveness remains stable despite this inherent randomness.

\textbf{External Validity.} Our evaluation focuses on three open-source DBMSs (DuckDB, PostgreSQL, SQLite) and two LLM families (Qwen and Llama), which may limit the generalizability to other systems. However, these DBMSs represent diverse architectural designs and cover a broad spectrum of SQL dialects. The consistent improvements across different model sizes (7B to 32B parameters) suggest that MIST is generalizable to similar systems and models.



\textbf{Construct Validity.} We use code coverage (line, function, and branch) as the primary metric for evaluating test effectiveness. While coverage is a widely accepted proxy for test quality, high coverage does not guarantee the absence of bugs or complete functional correctness. We acknowledge that our evaluation does not include explicit fault detection analysis, such as counting crashes or assertion failures triggered by the generated test cases. Nevertheless, the substantially higher coverage achieved by MIST suggests that it exercises deeper execution paths and edge cases that are more likely to expose latent faults compared to the baseline approach. Future work could evaluate fault-finding effectiveness through mutation testing or analysis of discovered bugs.

\section{Related Work}
\label{sec:related}
\subsection{LLM-based Testing}

The advent of LLMs has introduced new paradigms for automated software testing. Existing LLM-based testing approaches can be divided into two categories: direct generation and offline augmentation. For direct generation, Fuzz4All~\cite{fuzz4all} leverages LLM-driven prompt generation and mutation to create diverse test inputs for multiple programming languages. WhiteFox~\cite{YangDLY0J024} applies a two-model approach for white-box compiler fuzzing, where one LLM analyzes compiler source code and another generates targeted test programs. KernelGPT~\cite{YangZZ25} enhances OS kernel fuzzing by prompting LLMs to synthesize system call specifications from documentation. For offline augmentation, MetaMut~\cite{OuL0024} uses LLMs to automatically generate domain-specific mutators for compiler fuzzing, separating the costly LLM inference from high-throughput execution. Different from these approaches that focus on general-purpose programming languages, compilers, or operating systems, our work targets DBMS testing, where the challenge lies in adapting to proprietary SQL dialects and achieving high code coverage with lightweight local models under industrial resource constraints.

\subsection{Test Case Generation for DBMSs}

Existing methods for generating test cases for DBMSs can be divided into two categories: mutation-based and generation-based testing. Mutation-based approaches modify existing SQL statements to produce new test cases. Sedar~\cite{Fu0W024} improves input diversity by transferring SQL seeds across DBMS dialects. Griffin~\cite{FuLWWJ22} uses a grammar-free approach with a lightweight metadata graph to guide semantically-correct query mutations. Generation-based approaches construct queries from scratch using predefined rules. SQLsmith~\cite{SQLsmith} generates random yet well-formed SQL queries by leveraging schema metadata. SQLancer~\cite{SQLancer1,SQLancer2,SQLancer3} employs hand-crafted dialect-specific generators with sophisticated test oracles to detect logic bugs. Different from these traditional approaches that rely on manual grammar engineering or random mutation, our work is the first to leverage lightweight LLMs with hierarchical feature guidance and MCTS-based mutation guided by coverage feedback to systematically improve code coverage.

\section{Conclusion}
\label{sec:conclusions}
In this paper, we propose MIST, a framework for generating high-quality SQL test cases using lightweight LLMs. By combining feature-guided generation with error feedback and MCTS-based mutation guided by coverage feedback, MIST addresses the challenges of limited model adaptability and insufficient test diversity in industrial DBMS testing. Experiments on three widely-used DBMSs with four LLMs demonstrate that MIST greatly outperforms baseline approaches across line, function, and branch coverage, while achieving the highest coverage in critical modules such as the Optimizer. As a practical solution for resource-constrained industrial environments, MIST contributes to more reliable and thoroughly tested database systems. In future work, we plan to extend MIST to additional DBMSs and explore more sophisticated mutation strategies.


\normalem
\balance
\bibliographystyle{ACM-Reference-Format}
\bibliography{ref}

\end{document}